\documentclass[twoside]{article}

\usepackage{lipsum} 

\usepackage[sc]{mathpazo} 
\usepackage[T1]{fontenc} 
\usepackage{microtype} 

\usepackage[hmarginratio=1:1,top=32mm,columnsep=12pt,textwidth = 480pt ]{geometry} 
\usepackage{multicol} 
\usepackage[hang, small,labelfont=bf,up,textfont=it,up]{caption} 
\usepackage{booktabs} 
\usepackage{float} 
\usepackage{hyperref} 

\usepackage{lettrine} 
\usepackage{paralist} 

\usepackage{abstract} 

\usepackage{titlesec} 
\renewcommand\thesection{\Roman{section}} 
\renewcommand\thesubsection{\Roman{subsection}} 
\titleformat{\section}[block]{\large\scshape\centering}{\thesection.}{1em}{} 
\titleformat{\subsection}[block]{\large}{\thesubsection.}{1em}{} 
\usepackage{fancyhdr} 
\title{\vspace{-15mm}\fontsize{22pt}{10pt}\selectfont\textbf{The Compass for Statistical Researchers}} 

\author{
\large
\textsc{Daniele Durante$^{*}$, Davide Vidotto$^{\dagger}$ and Sabrina Vettori$^{\ddagger}$}\\[2mm] 
\normalsize Department of Statistical Sciences, University of Padova$^{*}$\\  \normalsize Department of Methodology and Statistics, Tilburg University$^{\dagger}$\\  \normalsize CEMSE Division, King Abdullah University of Science and Technology$^{\ddagger}$. \\ 
\normalsize \href{mailto:durante@stat.unipd.it}{durante@stat.unipd.it}$^{*}$,  \href{mailto:d.vidotto@tilburguniversity.edu}{d.vidotto@tilburguniversity.edu}$^{\dagger}$, \href{mailto:sabrina.vettori@kaust.edu.sa }{sabrina.vettori@kaust.edu.sa }$^{\ddagger}$
\vspace{-5mm}
}
\date{}

\begin{document}

\maketitle 

\thispagestyle{fancy} 


\begin{abstract}

We have hiked many miles alongside several professors as we traversed our statistical path -- a regime switching trail which changed direction following a class on the foundations of our discipline. As we play the game of research in that limbo between student and academic, one thing among Prof. Bernardi's teachings has never been more clear: to draw a route in the research map you not only need to know your destination, but you must also understand where you are and how you arrived there.

\end{abstract}


\begin{multicols}{2} 
\section*{Where are we?}

Students who are set in their choice to study a specific discipline since the beginning are like {\em outliers}. A few dots far away, determined and with clear ideas, which are likely to be generated from a latent process of decisions and education very different from that random mechanism matching the ambitions of the majority of students with their university route.

We are no exception. Our decision to study Statistics resulted from the minimization of different loss functions, grossly calibrated and solved with few data. Since then it has been a shared journey, made of several hikes in which we have filled that bag of skills and knowledge to carry with us on the journey towards the future. A regime switching path, which converged towards the PhD: a common destination of one journey and a starting point of a new one. 
 
Among all these {\em structural breaks} in our statistical journey, the one that occurred during a class on the foundations of Statistics was certainly one of the most decisive. An {\em educational revolution} leading us toward an inevitable paradigm shift, or better said, towards an ontological and epistemological enlargement of the {\em primitive} methodological paradigm we had prior to that class. We were already capable of {\em playing} with several, sometimes even sophisticated, statistical techniques. But had we asked ourselves if these phenomena existed (ontology)? And if they existed, were we sure it was possible to learn them (epistemology)? If yes, how (methodology)?

The answers to these questions are all ``no'', we had no idea about the response. Before studying the foundations of our discipline we were simply playing carefree in the Sciences' backyards, without wondering whether those gardens really existed or if statisticians were allowed to enter. We were like that conceited person who studies nature through statistical techniques, before making sure, as Galileo Galilei claims, that it is truly written in a mathematical language whose fonts are circles, squares and geometries.

Where were we? Was Statistics {\em ``dull, disreputable, prosaic and misleading [...], third rate discipline [...] a jackal picking over the bones and carcasses of the game that the big cats, the biologists, the physicists and the chemists, have brought down''}? Or rather {\em ``a wonderful discipline, including mathematics and philosophy, analysis and empiricism [...]. Which demands clear thinking, good judgement and flair [...], telling us how to turn data into decisions [...]. A Science dealing with the very essence of the universe: chance and contingency''}? Among these two definitions given by Stephen Senn in {\em Dicing with Death} (2003), we had hope in the second one to prove to be true. That foundations class confirmed our expectations, while adding  the further awareness that {\em ``data are not statistics as well as the marble is not the sculpture''}. A sentence from Bernardo Colombo which Lorenzo Bernardi commented in his contribution {\em Statistics and Mass Media} (2001), highlighting how  {\em ``both need to be modeled to gain intrinsic value and both somehow have the breath, originality and inspiration of their creator''}.

This is the story of our journey across the foundations of Statistics. From the epistemological consciousness, towards the methodological reasoning, until reaching the understanding of the techniques. Later, we will go beyond the path, placing Statistics in the map of Sciences, while looking at its final goal: the communication.

\section*{Back to the methodological future}
Prior to learning the foundations of Statistics, for us testing a statistical hypothesis was a simple procedure: simply subtracting a value from the mean, dividing it by a measure of variability, and checking whether the result was within a certain rejection region. We certainly had no idea that what we were doing was instead placed in a much wider methodological framework. Our immediate {\em declaration of war} was in fact only the final moment of a richer research path which required, among others,  a careful comparison with existing contributions, the formulation of specific hypotheses, the identification of the aggregate to study, and the choice of adequate analysis instruments. But why did we need this researcher's shopping list? Our calculations would still have remained the same.

Continuing our journey back to the future, the answer we gave to the previous question was ``yes", the results would have been the same, but without the {\em warranty} of the method they would have had zero weight in the scientific progress. We started looking at the method as the {\em insurance} of knowledge. An {\em ``ars bene disponendi seriem plurimarum cogitationum'' [art of ordering accurately multiple thoughts]} framed within specific historical coordinates, and with clear epistemological roots marking that invisible and sometimes very thin border between science and non-science. Where were we then?

In that ancient map representing the history of Science our coordinates were {\em inductive inference}, {\em experimental method}, {\em intersubjectivity of Science} and  {\em cross-fertilization}. A paradigmatic north-south-east-west where Statistics, intended as the Science aimed at balancing the empirical evidence with the hypotheses and conjectures of reason, was predestined to become the queen. We realized that our little calculations were sons of a new scientific method arising during the 16th and 17th centuries from a renewed confidence in the human abilities. A period of {\em Renaissance} that gradually broke the barrier between natural and artificial in favor of a re-evaluation of the technical and practical tools, and in which artists such as Brunelleschi, Mantegna and Leonardo da Vinci were also experts in urban planning, architecture and ballistics. A new worldview that abandoned the {\em Aristotelic} distinction between knowledge directed to practice and knowledge devoted to the contemplation of truth, favoring a much deeper experimental dialogue with the nature. 

Our little problems were thus the result of a new relationship with phenomena, no more characterized by a passive observation based on laws {\em ``in libris'' [in books]}, but an experimental one. The statistical techniques we used to process our data became instead like the telescope Galileo pointed skyward in 1609. Reliable research tools in prolonging the senses to favor a deeper understanding of nature. We discovered, finally, that our hypothesis testing procedure was actually part of a more ambitious attempt of {\em ``secare naturam'' [slicing the nature]}. We felt like little Galileos involved in a research process that recalls his study on falling bodies in 1608, where the comparison between each measured distance with what {\em ``should have been''} (on the basis of previous theories) represented the close relationship between theory and meaningful experience. We were freeing ourselves from the concept of {\em ``auctoritas'' [authority]} to venture into a growing education towards the critical meaning and the reasonable interrogation of nature within the guidelines of the inductive inference. 

But, again, why was the method so important? The answer is that it can't be any other way, according to a notion of Science which favored cross-fertilization and introduced the intersubjectivity as a form of objectivity. Citing Leonardo da Vinci {\em``those who love practice without theory are like the sailor who boards ship without a rudder and compass
and never knows where he may cast."} The method was the rudder and the compass guaranteeing internal consistency, transparency and reproducibility. Essential elements to allow comparison and monitoring of developments of knowledge in a slow process of progressive accumulations made by corrections and substitutions. An iceberg-shaped Science whose great discoveries and revolutions arise from a much broader layer of minor theories and findings, sometimes falsified and not visible to human eyes.

\section*{Hot data \& doctrines of uncertainty}

Data and probability are the statistician's daily bread. Citing Edward Deming, it was clear to us from the beginning that {\em ``without data you are just another person with an opinion.''} However, we didn't expect that these two basic ingredients in the kitchen of the statistician were indeed so elaborate and delicate. 

For us, data were initially {\em cold} numbers that someone gave us and which in turn we used to feed our statistical-probabilistic instruments to answer certain questions. During that foundations class, for the first time we started instead seeing them as {\em hot} information, and their collection process as a further expression of the artistic value of our discipline. Consider the Social Sciences, where the statistical survey became a slow artistic and painstaking process in which the statistician played cat-and-mouse with {\em ``chunks''} of information so as to obtain the maximum quality from the raw material harvested. It was necessary to present the purpose of the survey and then consider general questions before moving slowly towards increasingly specific aspects introducing the interviewee to the {\em ``threatening''} questions. A {\em first date with data}, where the statistician had to go through a series of ordered lovely actions, hoping that the social phenomenon of interest showed itself in all sincerity.

What about probability? We were able to play with several probabilistic tools. But from which remote planet was probability coming from? Or maybe we had learned it from ancient hieroglyphics? Was the concept universal? The first signboard along the fascinating trail that traces the historical and philosophical foundations of probability indicated it was coming from the planet {\em New Scientific Method}, which we had already explored. Citing Costantini's {\em Historical and Philosophical Foundations of Statistics and Probability} (2004), {\em ``the probability was the notion with which, during the breakthrough that led to the modern era, we tried to discover the laws of phenomena characterized by uncertain behavior.''}

Our Martians were therefore Laplace, von Mises, Jeffreys and De Finetti. The hieroglyphs were instead written on books like Laplace's {\em Th\`eorie analytique des probabilit\`es} (1812), Von Mises' {\em Wahrscheinlichkeit, Statistik und Wahrheit} (1928), De Finetti's {\em Sul significato soggettivo della probabilit\`a} (1931) and Jeffrey's {\em Theory of Probability} (1939). For us probability was no longer just a number between 0 and 1, which expressed a measure of {\em certainty} of an event, but took the shape of a {\em history about the beliefs in the uncertainty}. A {\em branching process} starting from the fact that nature is not is deterministic, and later branching out in different {\em doctrines} to define its uncertainty.

The way we initially viewed probability as the ratio of the favorable cases to the total number of cases was only one possible definition, the {\em classical} one formalized by Laplace. What about the others? We had unconsciously been using one of these definitions, the Von Mises' {\em frequentist} one, since our early classes in inference when we considered probability in terms of the limit of a relative frequency with which an event occurs in a very large (infinite) population. Having a frequentist background, we saw the other two, the De Finetti's {\em subjectivist} and Jeffrey's {\em logicist} definitions, as perhaps more surprising but no less interesting. The first placed the subject in the foreground, defining the probability as a degree of confidence that an individual assigns to the occurrence of an event according to his prior knowledge. The second instead defined probability as a logical relationship that exists between something known and something that is not, highlighting the objectivity (outside the subject) of the relationships. 

The question at this point was: why should we proceed in our journey with this {\em historical} baggage on probability? Simply because each of the paradigms of statistical inference arises from the aforementioned definitions of probability. We realized how the Fisherian approach inherited the {\em frequentist} concept of probability in exploiting information from the observed data (seen as realizations from a true unknown generative process) to appropriately estimate the generative mechanism and evaluate how sensitive this reconstruction was to the fact that observed data were only one sample of many possible. The frequency-decision theory adopted the same definition of probability, but changed the main focus of statistical inference to provide rules for action in situations of uncertainty. The subjective Bayesian approach was instead epistemologically and methodologically different in explicitly  studying how subjects' initial {\em confidence} (prior knowledge) about the uncertain generative mechanism  changed in the light of data. It was clear therefore how the {\em subjectivist} definition of probability was more suitable within this paradigm. The methodology remained mostly similar in the objectivist Bayesian approach, with an {\em epistemological} shift from prior knowledge to prior ignorance, progressively abandoning the subjective idea to reach the objective one stressed by the {\em logicist}  definition.

Far from the scope of providing a detailed and comparative overview of these paradigms (see Barnett's {\em Comparative Statistical Inference} (1999) for an extensive discussion), the main lesson we learned was that the choice of each paradigm of statistical inference carries specific definitions of probability we should be aware of in order to avoid confusion in choosing our techniques and in drawing our conclusions.

\section*{An ancillary science}
Now that we had found our coordinates in the map of Science and read the instructions of some {\em toys} we enjoyed playing with each day, we could fully understand a definition of Statistics often repeated by Lorenzo Bernardi. An {\em ancillary science}. A discipline that is fundamental to all other Sciences. The main actor in that slow process of donning with new certainties and doffing of falsified thesis that we had come to know by the name of {\em scientific progress}. Finally, not only did we appreciate Tukey's claim {\em ``the best thing about being a statistician is that you get to play in everyone's backyard,''} but we could also give a name to each of these backyards, understanding at the same time when we could enter and in which game we could play. We even discovered that these gardens had much more to offer than we initially thought.

There was the garden of Medicine, made for example of growth curves, odds, and survival models. In the Sports and Betting backyard we found sophisticated methods to model sports results; see also contributions in the third issue of volume 27 of {\em Chance}. The garden of Finance and Economy was covered with refined micro- and macro-econometric models and roller coasters of time series. At the end of the race the watchword was forecast. Certainly Sharpe, Miller and Markowitz in 1990 and Engle and Granger in 2003 must have enjoyed more than anyone else winning the Nobel Prize for economics. Prediction, especially of extreme events, was a key also in the backyard of Natural Sciences to provide tools for planning in agriculture and to anticipate natural disasters. In the garden of Industry we met, among others, experimental design and control charts for evaluating the conformity of the products with their required characteristics to avoid unpleasant surprises for customers. The latter became kings in the garden of Marketing, flooded with tons of data from the web which only data mining techniques could have been able to transform into useful information. Actors within certain social phenomena replaced customers in the park of Social Sciences, studded with social indicators, networks and many other sociometric instruments.

This was only one wing of a huge {\em Neverland} for statisticians, embracing all fields of knowledge such as physics, astronomy, chemistry, biology, psychology, archeology and many others. A {\em Disney World} that is constantly renewed to keep up with the times, new technologies and issues; where Statistics becomes Senn's (2003) {\em ``wonderful discipline that includes mathematics and philosophy, analysis and empiricism,''} and the new attractions are Neuroscience, Bioinformatics, Computer Science, Web Marketing, Sentimental Analysis and Applied Criminology.

An amusement park we wanted to celebrate in the {\em Statistical Calendar} (cal.stat.unipd.it/index.html) as well as in the video {\em My Statistician Friend} (www.youtube.com/watch?v=yU2qQywUnnU). Two dissemination projects perhaps naive, but passionate, which we would have never considered without having attended that foundations class.

\section*{The sexy job?}
{\em ``I keep saying the sexy job in the next ten years will be statisticians. People think I'm joking.''} At this point in our journey we have no doubts that the chief economist of Google Hal Varian was not joking. How can a discipline that combines art, science and technology not be sexy? However, the more we looked at the collective idea of Statistics, the more it seemed to be commonly perceived as Stephen Senn's (2003) {\em ``nasty old lady, you don't know her, but she loathes you already,''} or {\em ``like Australia, everyone knows where it is but no one wants to go there. Except that people do want to go to Australia.''}

Every time we found ourselves discussing Statistics among students from other fields, we noticed how after a few minutes they started surfing far away in a slow decaying process converging everywhere except in a neighborhood centered around {\em us} and with radius  {\em ``from here we do not listen to what you are speaking about.''} The few brave hearts who rarely remained started asking a series of questions randomly sampled from that set of conceptualizations and stereotypes, such as the famous {\em Trilussa's chicken} or newer versions like the one claiming that if you have your head in the oven and your feet in the freezer then your averaged body temperature is within the limits of healthy living.

In that complex {\em classification tree} grouping students according to their field of study, we found ourselves in an isolated branch representing the semi-known disciplines. This fact sounded extremely dissonant. If the goal of Statistics is to transform data into information, why wasn't this information being acknowledged? We wanted to understand why, after playing in the Sciences backyards, they didn't {\em come up and see our etchings}. The answer we gave to this question was that it wasn't enough to have {\em etchings} in our room, but it was also necessary to know how to correctly present them and how to properly invite Sciences.  

In that slow process of reconciliation with the other Sciences we understood the need to avoid {\em messy floods of data} and instead provide transparent information, favoring the opportunity to recognize our conceptual choices and report the general framework our analyses and interpretations came from. In parallel, it was important to create an {\em airlock} to balance the need to inform, sometimes with sophisticated methods of Statistics, with the aloof and suspicious attitude of some interlocutors who had to transform information into decisions. This armistice would have been possible only when the first would abandon its {\em arrogance} and, quoting George Box, the {\em ``bad habit of falling in love with its models,''} while the latter would have changed that attitude ranging between quantitative mythology and censorship.

How could we avoid the {\em arrogance}? Simply accepting the idea that many fields of statistical research are {\em relative}. Citing Bernardi (2001) {\em ``the spirit of statistical research can be explained by this story: to the friend who asks how's the wife, the economic statistician will reply: compared to when? while the social statistician will ask: with respect to whom?''} Part of our {\em arrogance} came from the need to always provide absolute answers resulting from naive conclusions, rather than proper exploitation of the analyses.

Our checklist ended with an {\em apology of simplicity}: we had to throw out the high-tech weapons and first learn to hit little birds with a slingshot instead of a cannon. Developing sophisticated statistical models was certainly much more stimulating than the slow process of verification of hypotheses and results somehow already present in the collective ideas. However, as Bernardi (2001) claims, {\em ``while the first is based on a scientific approach which sometimes proves precarious and naive, the second is always a solid starting point for new curiosities and fruitful intuitions.''} 

An exercise of {\em ``restating the obvious so as such remains,''} to train ourselves in the effective communication of finest methods, balancing sophistication and simplicity. Two not necessarily rival aspects, but difficult to reconcile for fear of trivializing the contribution of Statistics. Our {\em masterpieces} should not, therefore, have been merely an expression of {\em Pop artists}, but they also had to reflect the careful work of {\em copyists monks} engaged in that fundamental work of {\em ``translation and simplification of teaching, easily accessible for the reader.''}

Only by accepting this responsibility will we make peace with the other Sciences, allowing them to appreciate our  {\em etchings}. Conversely if we do not take proper care in communication, then we will remain that conceited guy admiring himself in the mirror without realizing he is indeed alone in a corner while all the Sciences are enjoying the dance floor of scientific progress.


\section*{Acknowledgments}
We are grateful to Giulio Peruzzi for his precious teachings on philosophy and history of Science. We also thank Bruno Scarpa, Antonio Canale, Maria Terres, Jacopo Soriano and Davide Salanitri for the fruitful discussion and edits on an early version of the manuscript.

\end{multicols}


\begin{thebibliography}{99}

\bibitem[{Barnett}{1999}]{barn_1999}
Barnett, V. (1999).
\newblock {\em Comparative Statistical Inference}.
\newblock John Wiley {\&} Sons, Inc.

\bibitem[{Bernardi}{2001}]{ber_2001}
Bernardi, L. (2001).
\newblock Statistica e mezzi di comunicazione di massa.
\newblock In Tuzzi, A., editor, {\em Dall'intervista alla notizia}, pages
  241--251. Edizioni Sapere.

\bibitem[{Costantini}{2004}]{cost_2004}
Costantini, D. (2004).
\newblock {\em I fondamenti storico-filosofici delle discipline
  statistico-probabilistiche}.
\newblock Bollati Boringhieri.

\bibitem[{De~Finetti}{1931}]{def_1937}
De~Finetti, B. (1931).
\newblock Sul significato soggettivo della probabilit\`a.
\newblock {\em Fundamenta Mathematicae}, 17:298--329.

\bibitem[{Jeffreys}{1939}]{jef_1939}
Jeffreys, H. (1939).
\newblock {\em Theory of Probability}.
\newblock The Clarendon Press.

\bibitem[{Laplace}{1812}]{lap_1812}
Laplace, P.~S. (1812).
\newblock {\em Th\`eorie analytique des probabilit\`es}.
\newblock Ve Courcier.

\bibitem[{Senn}{2003}]{sen_2003}
Senn, S. (2003).
\newblock {\em Dicing with Death}.
\newblock Cambridge University Press.

\bibitem[{Von~Mises}{1928}]{von_1928}
Von~Mises, R. (1928).
\newblock {\em Wahrscheinlichkeit, Statistik und Wahrheit}.
\newblock Von Julius Springer.

\end{thebibliography}
\end{document}